\documentstyle[prb,aps,psfig]{revtex}
\def\be{\begin{equation}}
\def\ee{\end{equation}}
\def\bc{\begin{center}}
\def\ec{\end{center}}
\begin{document}
\input epsf.sty
\twocolumn[\hsize\textwidth\columnwidth\hsize\csname %
@twocolumnfalse\endcsname
\draft
\widetext
\title{Phase diagrams of Ising films with competing interactions}
\author{W. Selke \$\ , M. Pleimling \dag\ and D. Catrein \ddag }
\address{
\$ Institut f\"ur Theoretische Physik, Technische Hochschule, D--52056 Aachen, Germany\\
\dag Institut f\"ur Theoretische Physik I, Universit\"at Erlangen-N\"urnberg,
D--91058 Erlangen, Germany\\
\ddag Institut f\"ur Stochastische Mathematik, Technische Hochschule, D--52056 Aachen, Germany }
\maketitle

\begin{abstract}
The axial next--nearest--neighbour Ising (ANNNI) model of
finite thickness is studied. Using mean--field theory, Monte Carlo
simulations, and low--temperature analyses, phase diagrams
are determined, with a distinct phase diagram for each 
film thickness. The robustness of the phase diagrams
against varying the couplings in the surface layers
is analysed.
\end{abstract}


\phantom{.}
]
\narrowtext
 
\section{Introduction}
In recent years magnetism in thin films of a few atomic
layers has attracted much interest, both
theoretically and experimentally \cite{Poul}. However, studies on
the influence of the layer thickness, $L$, on spatially modulated
magnetic structures seem to be very scarce \cite{Cow,Sut}, albeit the possible
lack of compatibility between the film geometry and
the modulations in the bulk as well as
the effect of the surfaces on the ordering phenomena
may lead to interesting features.   

In this article, we
shall deal with this topic by analysing phase diagrams
of the axial next--nearest--neighbour Ising (ANNNI)
model \cite{Selke,Yeo,Pleim} on a simple
cubic lattice. Due to its competing interactions, the model
displays, in the limit of infinitely many
layers, a phase diagram with a plenitude of commensurate
phases, including those springing from
the multiphase point at zero temperature and
those emerging from structure combination
branching processes at finite temperatures, as well as incommensurate
phases and a Lifshitz point. Many of these aspects
have been observed experimentally, in particular
in magnets, alloys, polymers, and ferroelectrics, see the
reviews \cite{Selke,Yeo,Pleim} and related recent
work \cite{Col,Schob,Tan,Mart,Schwahn}.

The intriguing structural complexities are severely affected
when the lattice consists of rather few layers perpendicular
to the axis of competition. Of course, in the case of periodic
boundary conditions for the top and bottom
layers, the bulk phases which fit to the
film thickness still exist \cite{Selke,Yeo,Pleim,Waa}. In the case of
free boundary conditions for the surface layers, describing more
realistically experimental
situations, novel surface--induced features may evolve. The aim
of this paper is to identify similarities and typical differences between
the phase diagrams in the limit of an infinite lattice
and for thin films with free boundaries.

To study the influence of the surfaces in more detail, we also 
varied the intralayer
couplings in the surfaces compared to those in the bulk. In semi--infinite
systems, one then encounters the well--known surface critical
phenomena with either ordinary or extraordinary and surface
phase transitions \cite{Bind,Diehl}, depending
on the strength of the surface and bulk couplings. Note that critical
properties of the surface magnetization at the Lifshitz point in the
semi--infinite ANNNI model have been analysed recently, refining
results of mean--field theory \cite{Bind2} by doing Monte Carlo
simulations \cite{Pleim2}. 

In the following, we shall mostly deal with thin films of
up to ten layers. For each
film thickness, distinct phase
diagrams in the (temperature--competition strength)-plane are
determined, using mean--field theory, low--temperature expansions,
and Monte Carlo simulations. Mean--field theory is found to usually provide
reliable guidance to the correct phase diagrams; qualitative shortcomings
are observed especially in the case of vanishing surface couplings. Brief
accounts of some of our findings for equal
surface and bulk couplings have been given before \cite{Sel2,Sel3}. 

The article is organized as follows: first the
model and the methods are presented, then the resulting
phase diagrams are discussed. The paper is concluded by a summary.

\section{Model and methods}

We consider the ANNNI model on a simple cubic lattice (setting the
lattice constant equal to one) for films of $L$ layers, $L > 2$, with
free boundary conditions for the surface
layers. Each spin $S_j$, at site $j$, can take only two values, $S_j= 1$
(spin 'up') or $S_j= -1$ (spin 'down'). The 
interactions are supposed to be ferromagnetic between
neighbouring spins in each layer, $J_0 \geq 0$, as well as between
neighbouring spins in adjacent layers, $J_1>0$, and to be
antiferromagnetic, $J_2<0$, between
axial next--nearest neighbour spins, distinguishing one of the three
cubic axes, say, the $z$--axis. The strength of the competition
between the interactions along the $z$--axis is $\kappa= -J_2/J_1$. The
intralayer couplings $J_0$ may be different in the (top and
bottom) surface layers, denoted by $J_s$, compared to those in
the bulk layers, $J_b$, with the ratio $r= J_s/J_b$. For
simplicity, we set $J_b= J_1$. Changing the interlayer coupling $J_1$
relative to the bulk intralayer interaction, $J_b$, modifies
only quantitatively the phase diagram in the
limit $L \longrightarrow \infty$ \cite{Salinas,Nakani,Rotthaus}.

The ground state properties of the ANNNI films are readily
obtained. We first consider $J_s > 0$. The 
spins in all layers are aligned
ferromagnetically at $\kappa < 1/2$. For $\kappa > 1/2$, and
even film thickness $L$, the ground state consists of pairs
of layers with, say 'up' spins, denoted as a
'2-band' \cite{Fisher1,Fisher2}, followed by a 2--band
of 'down' spins, being obviously two--fold degenerate by interchanging
the $+$ and $-$ spins. If
$L$ is odd, then for $1 > \kappa >1/2$, the ground state
comprises one 3--band and $(L-3)/2$ 2--bands, being $(L-1)$--fold
degenerate. For $\kappa > 1$, structures with one 1-band at the surface
and 2-bands are stable, with a 4--fold degeneracy. Both for even and
odd number of layers $L$, the degeneracy $D_L$ of the
multiphase point  at $\kappa= 1/2$ is given by a Fibonacci
sequence $D_L= D_{L-1} + D_{L-2}$, where $D_2= D_3= 2$, corresponding to
all possible combinations of $k-$bands with $k > 1$, as discussed
before \cite{Fisher2,Redner}. For
$L$ odd, at $\kappa= 1$,  $L+ 3$ structures have the
same ground state energy. For $J_s= 0$, the situation at $\kappa \geq 1$
is of special interest, $L$ being odd. At $\kappa = 1$, structures
with 2-bands followed by an arbitrarily ordered surface layer are
degenerate, due to the compensation of the interlayer
interactions between the surface spins and those in the
adjacent and next--nearest
layers. In the case of $L=3$ and $\kappa > 1$, the 
orientations of the spins in one of the surface layers are
random, and completey antiparallel to that random pattern in the other
surface; in the center layer, all spins have the same sign.

At non--zero temperatures, the stable phases have been determined using
mean--field theory, low--temperature expansions, especially about
the special points ($T= 0$, $\kappa= 1/2$) and
($T= 0, \kappa= 1$), as well as Monte Carlo
techniques.

In mean--field theory, the magnetization per
layer, $m_i$, $i= 1$,2,...$L$, in thermal equilibrium follows
from the standard equations \cite{Salinas,Jan,Bak,Dux}        

\begin{eqnarray}
 m_i = \tanh ((4J_0(i)m_i+ J_1(m_{i-1}+ m_{i+1}) \nonumber \\
&& \hspace*{-3cm}  + J_2(m_{i-2}+ m_{i+2}))/k_BT)
\end{eqnarray}
where $J_0(i)=J_s$ for $i= 1, L$ and $J_0(i)= J_b$ for the other
layers. Free boundary conditions are implemented by setting $m_i= 0$ for
$i=-1, 0, L+1, L+2$. To obtain the thermally stable magnetization pattern, we
solved Eq. (1) iteratively starting from all
possible combinations of fully ordered, $m_i= \pm 1$, or
completely disordered layers, $m_i= 0$, i.e. from, in principle, $3^L$
distinct configurations (the number may be reduced using symmetry
considerations). We then determined among the solutions the one
with the lowest free energy. Phase boundaries are identified by singularities
in the free energy and specific heat. Obviously, a fine scan of
the (temperature $k_BT/J_1$, competition strength $\kappa$)--plane
is rather computer--time consuming, and the full analysis
of the mean--field theory was done for films of up to $L=10$
layers. The ratio $r= J_s/J_b$ varied from 0 to 1.5, which
would cover, in the limit $L \longrightarrow \infty$, both
ordinary and  surface transitions.

The transition to the paramagnetic phase may be studied for films
of larger thickness, $L$, by analysing the linearized form of the
mean--field theory. From Eq. (1) one obtains a matrix of rank $L$, with
the eigenvalues determining the phase transition temperature, $T_c$,
and the eigenvectors describing the critical magnetization pattern. Films
of thickness up to $L= 50$ were considered, especially for
competition ratios close to that of the Lifshitz point in
the infinite system, i. e. $\kappa= 0.25$.   

The phase diagram close to the special ground states at $\kappa= 1/2$
and 1 may be investigated by using exact low--temperature
expansions \cite{Fisher1,Fisher2}. The stable phases
are identified
by calculating the free energy resulting from
spin excitations for all structures being degenerate
at the two special points. Indeed, to establish the
stable phases springing from the special points for thin films (we
studied films with up to ten layers), it usually suffices to do expansions
up to first order, involving merely a single spin flip.

Complementary to the low--temperature analysis, Monte Carlo simulations
may be applied to study the phase diagrams at higher temperatures. We
used both the standard single--flip Metropolis algorithm \cite{Binder2}
and a cluster--flip algorithm \cite{Pleim3} (attention may
be drawn to another cluster--flip algorithm to simulate
spatially modulated structures in Ising models \cite{Matsu}). The main aim of
our simulations has been to
check results of mean--field theory, restricting ourselves to selected
cases. Of course, the layers are now
finite, consisting of, say, $M^2$ spins. Thereby, finite--size effects
have to be taken into account (here, attention may be drawn to early
Monte Carlo work on the nearest neighbour Ising model on
thin films \cite{Binder0}). Typically $M$ was varied from 10 to 100. To
identify the structures and boundary lines of the
various phases, we computed the energy, the specific heat, the
layer magnetizations and corresponding histograms as well as
correlation functions between spins in different layers. Each run
was performed with about $10^6$ Monte Carlo steps per spin, when
using the single--spin--flip algorithm. In the cluster--algorithm,
in each run usually about $2 \cdot 10^5$ clusters were generated.

\section{Results}

For each film thickness $L$, distinct phase diagrams are
obtained. Before presenting them, we shall outline
some general features, as inferred from calculations
for films with up to ten layers.

One may distinguish between phases corresponding to ground states
and those which are stabilized only at higher
temperatures. Especially, the
ferromagnetic phase as well as phases consisting of 2--bands
augmented, for $L$ odd, by
one 3--band or one 1--band, evolve from the ground states. A
'$k$--band', at $T > 0$, means that spins
in $k$ adjacent layers are oriented predominantly in one
direction, preceded and followed by layers where the
spins are oriented predominantly in the
opposite direction. The 3--band tends to be at or near to the center
of the film at low temperatures, when it gains
entropy from easy spin flips in the two border layers of
the 3--band. The band, however, sticks to one of the surfaces
of the film when $J_s$ is rather weak so that entropy is
gained from energetically easy flips of surface spins, as may be readily seen
from low--temperature expansions.

Phases corresponding to ground states with 2--bands
and one 3--band at or near the center (or at the surface) of the
film may be stabilized only at higher temperatures, due
to entropic or symmetrization reasons, while at low
temperatures the 3--band sticks to the surface (or center) of the film.

The phases with one 3--band and one 1-band, $L$ odd, are
separated, for $J_s > 0$, by the transition line arising
from ($T=0, \kappa=1$); the line is at low temperatures of
first order. If the surface couplings $J_s$ are not very strong, the
line terminates in a critical point, above which the two phases
become indistinguishable. That point moves down to zero temperature
as $J_s$ vanishes.

Other stable low--temperature phases may spring from the multiphase
point ($T=0, \kappa= 1/2$) for $L> 5$. In fact, in the limit
$L \longrightarrow \infty$, a sequence of infinitely many periodic
$\langle 32^i \rangle$ phases, $i=$0,1,2,...,$\infty$, arises
from that point (here and in the following, we are using or 
adapting the standard notation \cite{Fisher1,Fisher2}, setting
the sequence of bands in a phase in '$\langle  \rangle$'--brackets). For thin
films, the low--temperature expansions indicate the
following systematics: If $L$ is a multiple of 3, the
$\langle 3^{L/3} \rangle$ structures become stable next to
the ferromagnetic phase; for $L= 3n +1 ( =3n +2)$, $n > 1$,
they are replaced by the
$\langle 3^{n-1}4 \rangle $ ($\langle 3^{n-1}5 \rangle $)
structures. Further
phases arising from the multiphase point show up
for $L > 7$, having, for $L= 8$, two 3--bands and
two 2--bands, and two 3--bands and three 2--bands for $L= 10$. The
3--bands stick to the surfaces of the film, when $J_s$ is
sufficiently weak, as before. An increasing number of
similar phases comprising 2--bands and
3--bands are expected to occur for thicker films. In comparison to
the $\langle 32^i \rangle$ phases, arising
from the multiphase point in the limit $L \longrightarrow \infty$, the
sequences of 2--bands and 3--bands are slighty rearranged
and modified, reflecting the influence of the surfaces.

The transitions between
the phases emerging from the multiphase point are, at
low temperatures, of first order. Note that, for films, some, but
not all these phases extend up to the transition line
to the paramagnetic phase, $T_c$.

Novel phases may become stable close to $T_c$. As
may be obtained readily from the linearized mean--field
theory, confirmed
by Monte Carlo simulations, the ordered phases directly below
this transition line are alternatingly, as $\kappa$ increases, symmetric
and antisymmetric with respect to the center plane (being obviously
a real layer for $L$ odd, and a fictitious plane
for $L$ even), changing thereby
the parity. Each change is associated with an abrupt decrease in
the average wavelength, as follows from a Fourier analysis of the
magnetization pattern, $m_i$. For odd $L$, the magnetization in
the center layer vanishes in the antisymmetric cases, denoted
as '$1_0$--band'. There, the
interlayer couplings with the neighbouring layers tend to
compensate. Such 'partially
disordered phases' are, of course not stable below the
ordering temperature of the two--dimensional Ising plane. They
do not exist in the limit $L \longrightarrow \infty$ \cite{Nakani,Rotthaus}.-- Of course, as
usual, the location of $T_c$ is largely overestimated in mean--field
theory, as seen from comparison with simulational data.

An interesting
feature of the transition line to the paramagnetic phase
is the point where the ferromagnetic structure becomes
unstable against spatially modulated structures, i.e. the
Lifshitz point in the limit $L \longrightarrow \infty$. In mean--field
theory, for rather thick films of up to 50 layers,
the surface magnetization is found to show intriguing crossover
effects with an effective critical exponent close to 1 in the
vicinity of that point, in agreement with
the mean--field analysis
on the semi--infinite ANNNI model \cite{Bind2}. In a recent
Monte Carlo simulation, the exponent has been shown to be
significantly affected by critical fluctuations, acquiring
a value of about 0.62 \cite{Pleim2}.    
   
Further phases
may show up at intermediate temperatures, in particular, the
$\langle 54 \rangle $ and $\langle 64 \rangle $ structues
for $L=9$ and 10. They are reminiscent
of the plenitude of commensurate phases
on the ferromagnetic side of the ANNNI model in the thermodynamic
limit \cite{Dux}, and more of them are expected to occur in
thicker films. Likewise, wider films are needed to
see indications of the systematic branching processes involving
structures with 2--bands and 3--bands, as occurs for
$L \longrightarrow \infty$ \cite{Dux}.

Note that the magnetization in a given phase may change sign in
some layers when varying temperature and/or strength
of couplings, keeping the symmetry of the
magnetization pattern about the center
plane (symmetric, antisymmetric or asymmetric).
For instance, for an (anti--)symmetric phase such a reversal
of the sign may happen in the magnetization of the surface
layers belonging originally to $k-$bands with $k > 2$, when $\kappa$ is
increased, especially for weak surface couplings. In 
any event, we
characterise a phase by the sequence of bands which
occurs at lowest temperatures in case of equal
surface and bulk couplings, following the unambiguous notation for
the infinite lattice \cite{Fisher2}.

We shall now discuss results for each film thickness, from
$L=3$ to $L=10$, illustrating the general features in selected cases
in more detail. 

In films with three layers, $L= 3$, the
ferromagnetic, the $\langle 21 \rangle $ and
the $\langle 11_01 \rangle$ phases show up. As
found in simulations, for
equal surface $J_s$ and bulk $J_b$ couplings, the ferromagnetic
phase is stable at $\kappa <1$, the other two phases occur at
$\kappa >1$, with the transition between the
$\langle 21 \rangle $ and the partially disordered phases being of
second order, see Fig. 1, and belonging to the universality class
of the two--dimensional Ising model. At $\kappa =1$, the
ferromagnetic and $\langle 21 \rangle $ phases coexist, with the
transition being of first order. By weakening the surface
interactions, $J_s$, the ferromagnetic phase extends to
larger values of $\kappa$, due to the entropic effect 
in the surface layers. In the limit
$J_s= 0$, the $\langle 21 \rangle $ and $\langle 11_01 \rangle $ phases
are completely suppressed (this aspect is not correctly described
in mean--field theory). On the other hand, by
enhancing the surface couplings, the
$\langle 21 \rangle $ and the $\langle 11_01 \rangle $ phases are 
favoured and stabilized already at
$\kappa < 1$, with the transitions to the ferromagnetic phase
being always of first order.   

\begin{figure}
\centerline{\psfig{figure=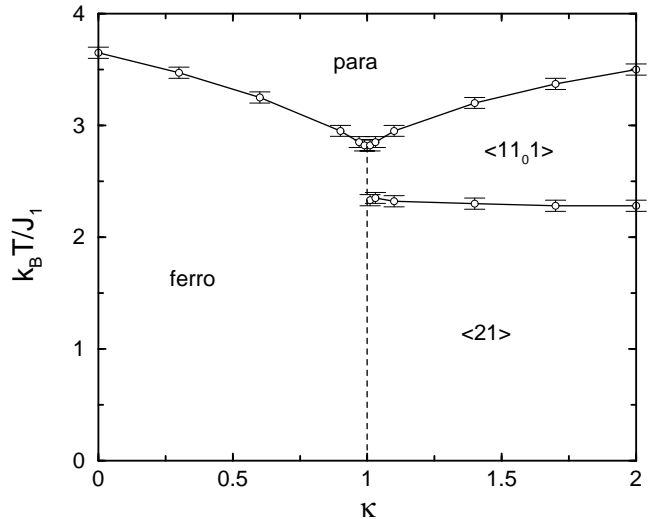,angle=270,width=3.3in}}  
\caption{Phase diagram of the ANNNI model for $L= 3$ and $J_s= J_b$ using
Monte Carlo simulations. The curves are guides to the eye. Phase
boundaries of first (second) order are denoted by dashed (solid) lines.}
\label{fig1}
\end{figure}

For $L= 4$, the phase diagrams are very simple, with two
ordered phases, the symmetric, ferromagnetic and the 
antisymmetric $\langle 2^2 \rangle$ phases being
separated by a first order transition line arising from the
special point $(T=0, \kappa= 1/2)$. By weakening $J_s$, the
range of stability of the symmetric phase is
again enlargened. The $\langle 2^2 \rangle $ structure
is obviously favoured when the surface layers are more
ordered, and thence its range of stability expands when
$J_s$ gets stronger.      

Somewhat richer phase diagrams are encountered when $L= 5$. For
$J_s= J_b$, one finds, increasing the competition ratio $\kappa$, at
low temperatures the ferromagnetic, the $\langle 32 \rangle $, and
the $\langle 2^21 \rangle $
phases, with the latter two becoming unstable against
the $\langle 21_02 \rangle $ and
$\langle 131 \rangle $ phases at higher temperatures. Phase
diagrams, obtained from mean--field theory and 
Monte Carlo simulations have been depicted before for
the case $J_s= J_b$ \cite{Sel2}. The transition
line between the two asymmetric (with respect to
the center plane) $\langle 32 \rangle $ and
$\langle 2^21 \rangle $ phases ends in
a critical point above which the phases become essentially
indistinguishable. By weakening $J_s$, the critical point moves to
lower temperatures. Eventually, in the limit of zero surface
couplings, there is only one asymmetric phase, with one of the
surface magnetizations going to zero on approach to
$(T= 0, \kappa= 1)$. The
phase diagram, as determined from simulations, is shown
in Fig. 2a. Note that the $\langle 131 \rangle $ structure, being
present close to $T_c$ in the mean-field
calculations, is squeezed out (as we checked
for $\kappa$ as large as 15). In turn, the range of stability of the
$\langle 21_02 \rangle $ phase is underestimated in
mean--field theory, similar to the situation for equal surface and bulk
couplings \cite{Sel2}. The
magnetic disorder in the center layer sets in roughly at the
transition temperature of the two--dimensional Ising model, because
then the interlayer couplings $J_1$ and $J_2$ to
the axial spins in the two adjacent and two next--nearest layers are
largely cancelled. When increasing the surface couplings
$J_s$, the $\langle 131 \rangle $ phase is stabilized, as shown in
Fig 2b, in qualitative agreement with
mean--field theory. Actually, the magnetization then tends to
be close to zero in the second and fourth layers, being rather
large with the same sign in the surface layers and with the opposite sign
in the center layer.  

\begin{figure}
\centerline{\psfig{figure=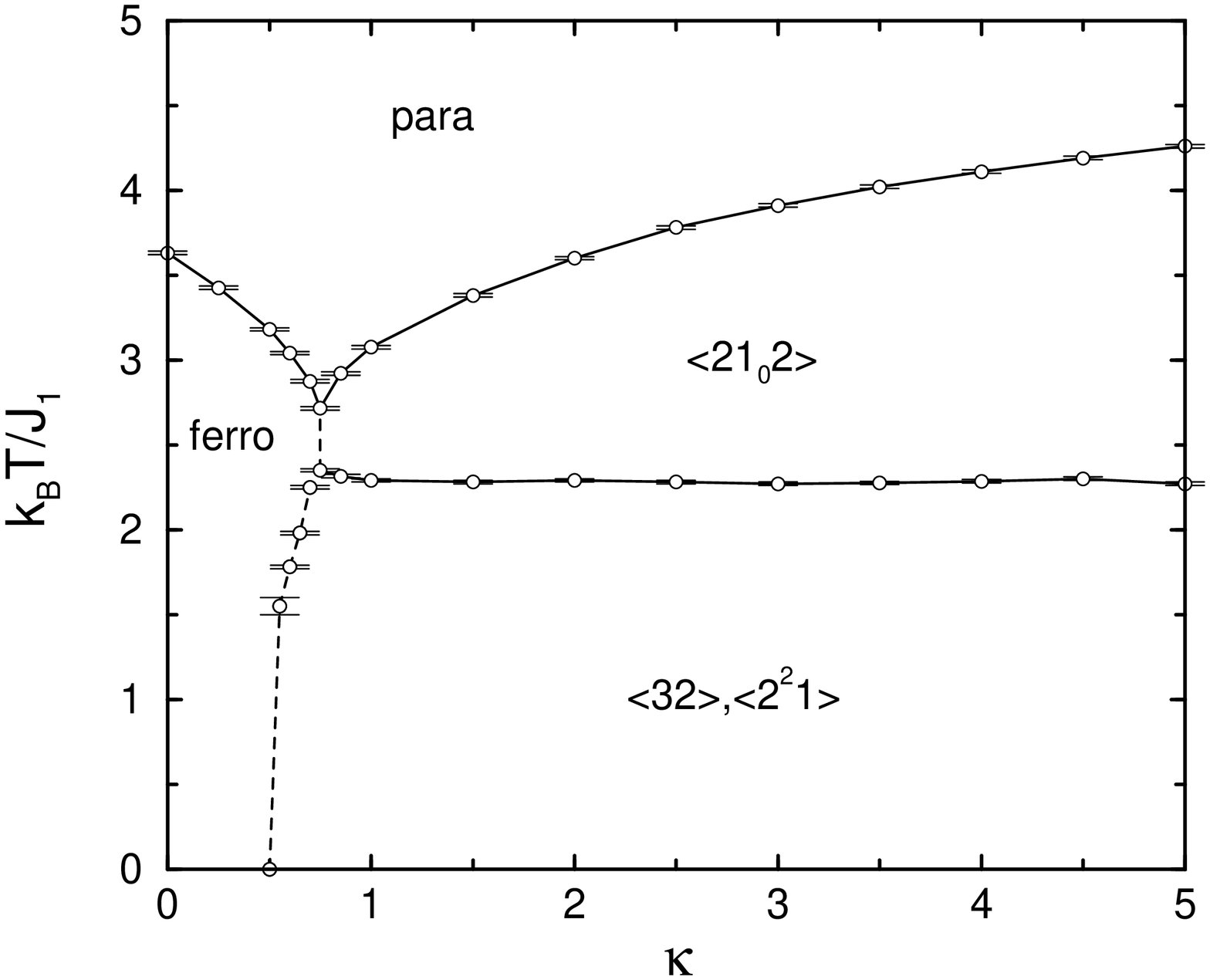,angle=270,width=3.3in}}  
\centerline{\psfig{figure=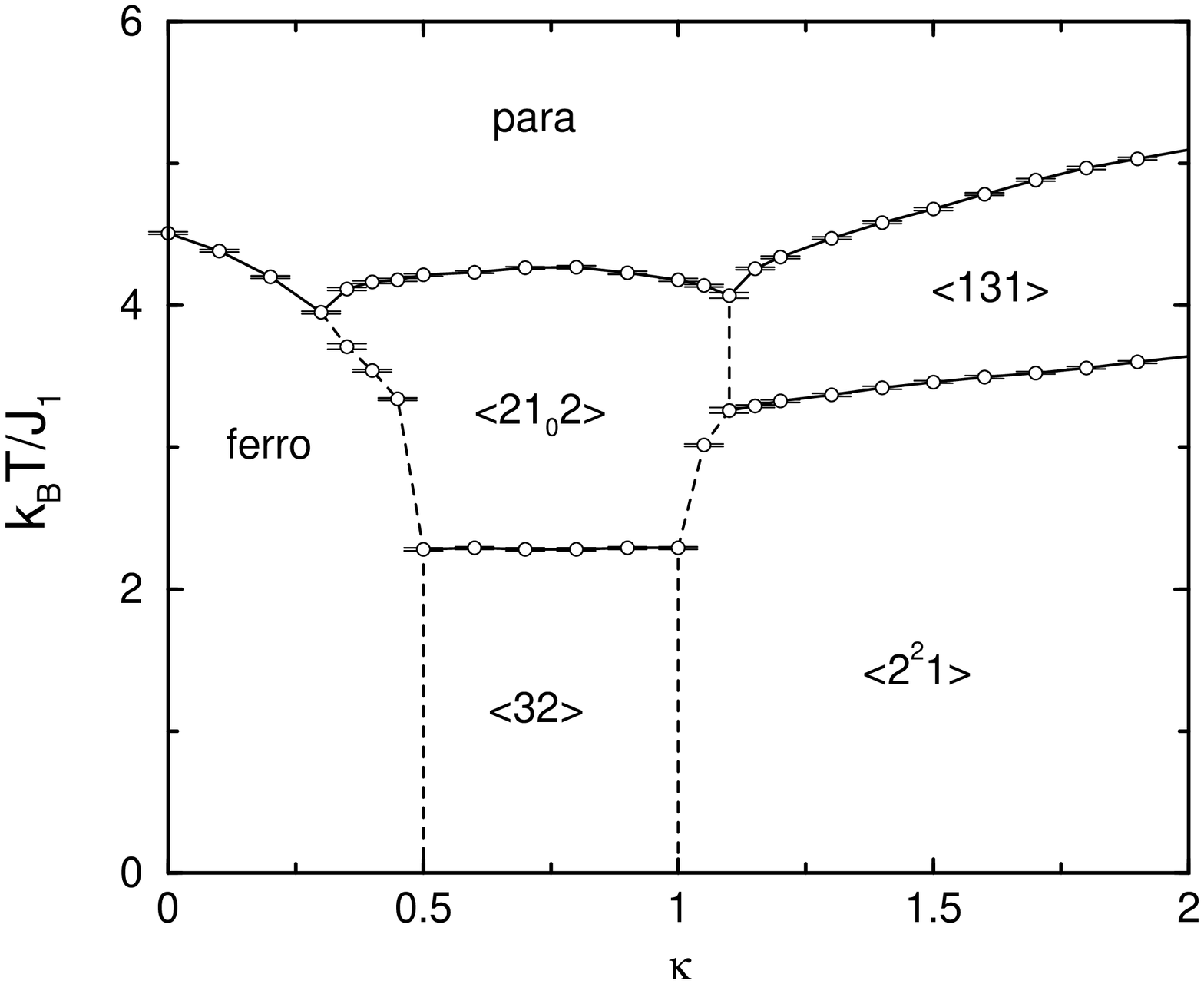,angle=270,width=3.3in}}
\caption{Simulated phase diagram of the ANNNI model
for films with five layers at $J_s/J_b$= (a) 0 and (b) 1.5.}
\label{fig2}
\end{figure}

For $L=6$, only three ordered structures exist, the ferromagnetic, the
$\langle 3^2 \rangle $, and the $\langle 2^3 \rangle $ phases. The
transitions between them are
of first order. By decreasing the surface interactions, the range
of stability of the $\langle 3^2 \rangle $ phase gets wider, with
1-bands at the surfaces for large competition ratio $\kappa$. This range
and that of the ferromagnetic phase shrink when enhancing the surface
couplings, and the $\langle 2^3 \rangle $ structure is
favoured. Predictions of
mean--field theory were qualitatively confirmed in simulations.

Adding another layer to the film, $L= 7$, leads to quite
different phase diagrams, with possibly seven distinct ordered
phases, namely the
ferromagnetic, $\langle 43 \rangle$, $\langle 31_03 \rangle$,
$\langle 232 \rangle$, $\langle 32^2 \rangle$,
$\langle 2^31 \rangle$, and $\langle 121_021 \rangle$ phases, as found in
mean--field theory. For $J_s= J_b$, the $\langle 232 \rangle$ phase
extends down to zero temperature at $1/2 < \kappa <1$, while
the $\langle 32^2 \rangle$ structure is stabilized only at intermediate
temperatures, where the entropic
effect of the small surface magnetization plays the
crucial role \cite{Sel2}. Examples of phase diagrams for different surface
and bulk couplings are depicted in 
Figs. 3a and 3b, for $r= J_s/J_b=$ 0.25 and 1.5, as obtained from
mean--field theory. Compared to the case of $J_s= J_b$,
the enhancement or weakening of
the surface couplings may lead to similar trends as for $L= 5$, but
also to novel features. As before, when
$J_s$= 0, there is no transition
line arising from the special point $(T=0, \kappa= 1)$, i.e. the
two asymmetric $\langle 32^2 \rangle$ and $\langle 2^31 \rangle$ structures
transform gradually into each other
at non--zero temperatures, with a vanishing magnetization in one
of the surface layers on approach to the special point. Indeed, even
the transition line to the neighbouring asymmetric $\langle 43 \rangle$ phase
terminates in a critical point, above which all these asymmetric structures
belong to the same phase. Perhaps somewhat unexpectedly, at 
$1/2 < \kappa < 1$, the $\langle 2^23 \rangle$ structure forms the
low--temperature phase when $r < (3 + \kappa)/4$, while
for stronger surface couplings the symmetric $\langle 232 \rangle$ phase is
stable down to zero temperature. In the former case, the symmetric
phase is stable next to $T_c$, see Fig. 3a. A similar entropy--driven
shift of the 3--band from the interior of the film to one of its
surfaces occurs at wider films, $L$ odd, as well, for sufficiently
weak $J_s$. The effect follows from the low--temperature analysis, and
it is described correctly by mean--field theory.  

In the case of $L=8$, six ordered phases are observed; the
phase diagram as obtained from mean--field theory
for equal couplings $J_s= J_b$ has been shown
before \cite{Sel2}. Apart from
the trivial ferromagnetic and $\langle 2^4 \rangle$ phases, there is
the $\langle 3^22 \rangle$ phase, which may, at sufficiently strong surface
couplings $J_s$, arise from the multiphase
point $(T=0, \kappa=1/2)$, with the
symmetric $\langle 323 \rangle$ structure, formed at higher
temperatures, remaining stable up to the transition line to the
paramagnetic phase, $T_c$. The latter phase is stable down to the
multiphase point for sufficiently weak surface
couplings, where entropy is gained, again, from 3--bands at
the surfaces, as follows from low--temperature
considerations, in agreement with mean--field
calculations. Actually, for small $J_s$, the
$\langle 323 \rangle$ phase persists
even for quite large values of the competition
ratio $\kappa$. Furthermore, the $\langle 53 \rangle$ phase may spring
from the multiphase point. The antisymmetric
$\langle 4^2 \rangle$ phase is stable close to $T_c$, extending down to lower
temperatures with decreasing surface couplings.

\begin{figure}
\centerline{\psfig{figure=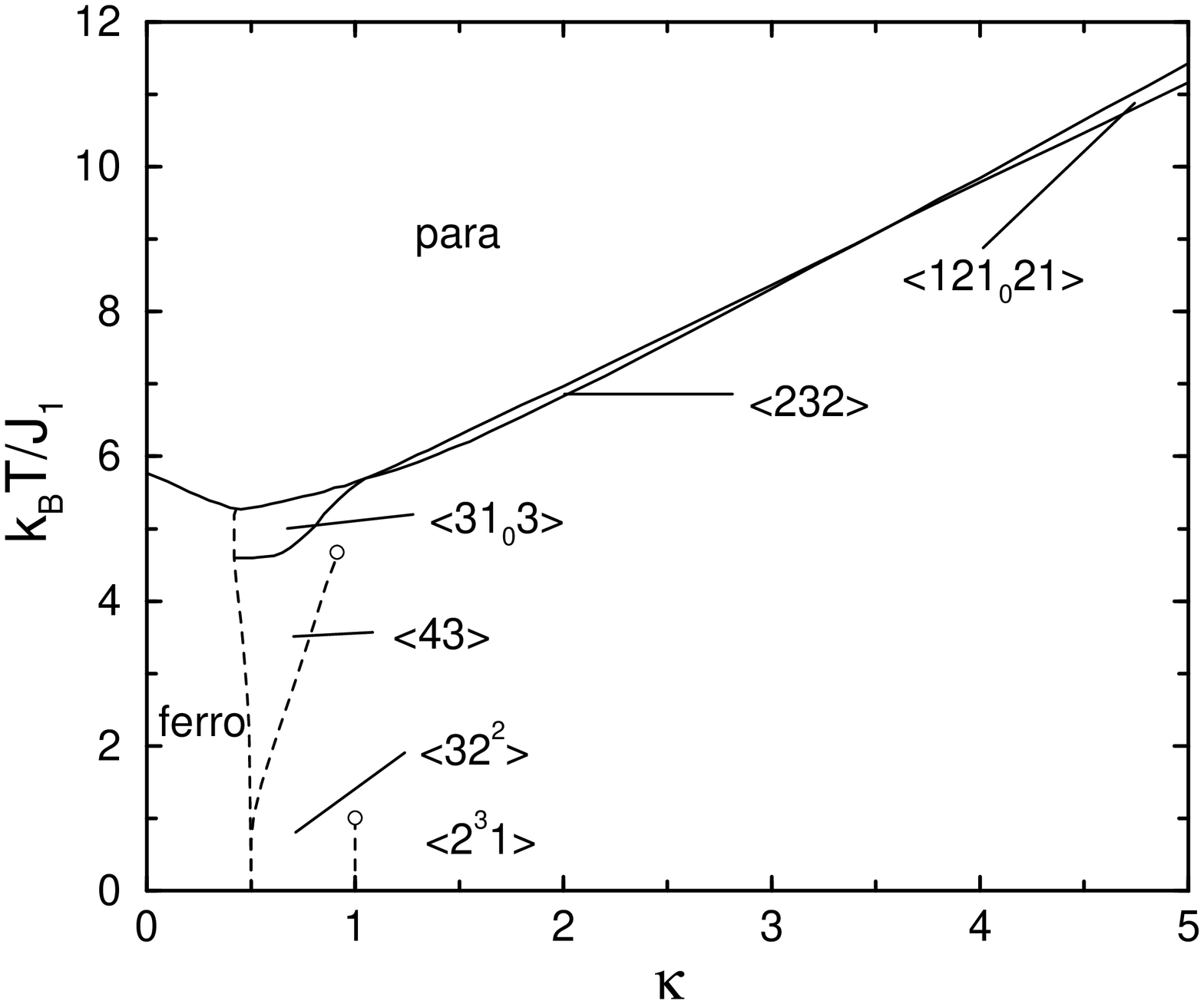,angle=270,width=3.3in}}  
\centerline{\psfig{figure=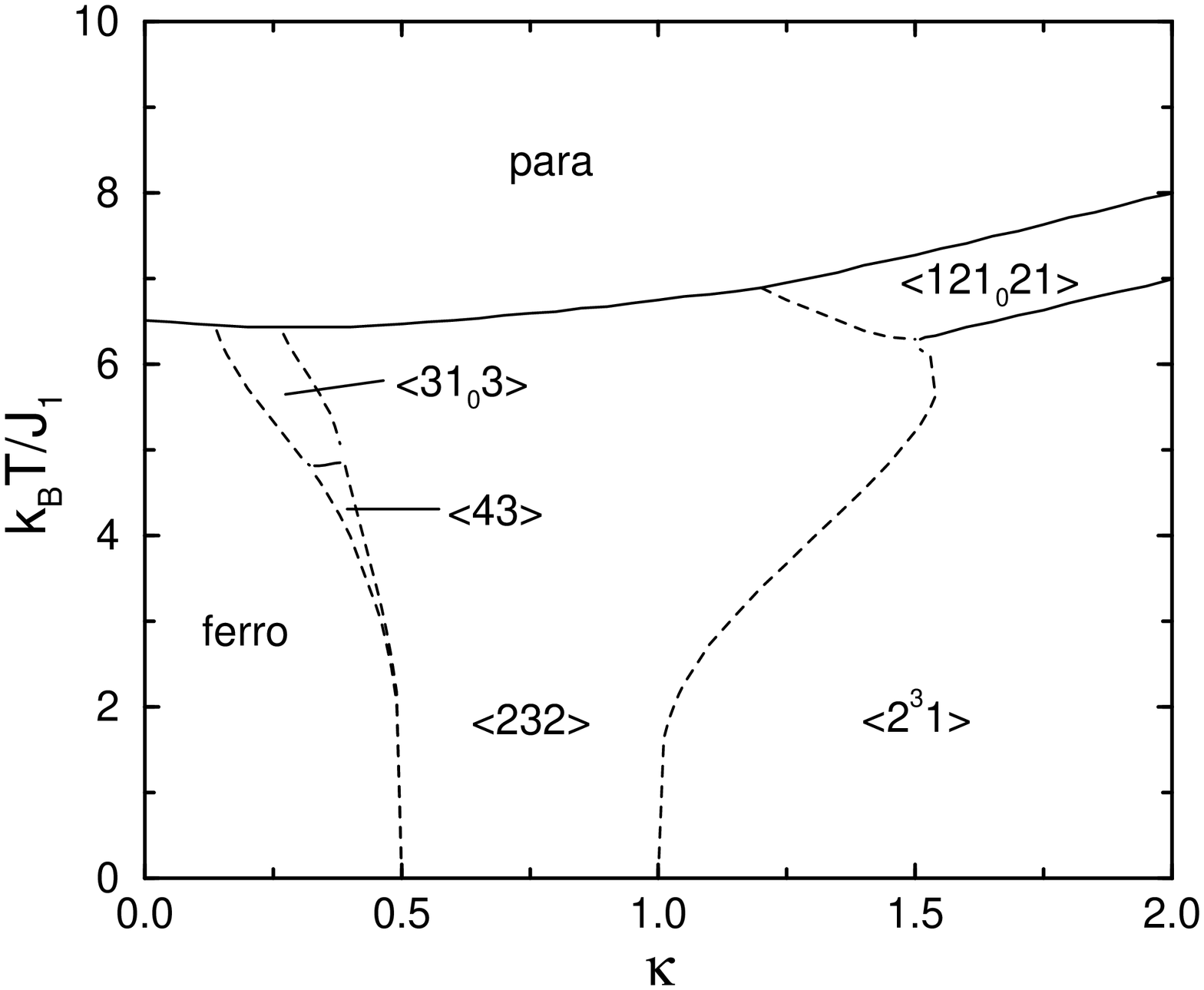,angle=270,width=3.3in}}
\caption{Phase diagram of the ANNNI model for $L= 7$ at
$J_s/J_b$= (a) 0.25 and (b) 1.5, as obtained from mean--field theory. The
critical point at the end of a line of transitions of first order
is denoted by an open circle.}
\label{fig3}
\end{figure}

In the case of nine layers, the rather complex phase diagrams
consist of up to nine ordered phases. Close to $T_c$, increasing
$\kappa$, the alternatingly symmetric and antisymmetric phases
are the ferromagnetic, $\langle 41_04 \rangle$, $\langle 3^3 \rangle$, 
$\langle 2^21_02^2 \rangle$, and
$\langle 12321 \rangle$ phases. At low temperatures, one
may, in addition, encounter, the $\langle 2^41 \rangle$ phase
as well as, for
fairly weak surface couplings, $r < (3 + \kappa)/4$, the
$\langle 32^3 \rangle$ phase, and, for stronger couplings, the
$\langle 2^232 \rangle$ phase at $1/2 < \kappa <1$, similar to the 
situation in the case $L= 7$. Again, this
feature follows from the low--temperature analysis, and it
is described correctly by mean--field theory, see
Fig. 4 (the case $J_s= J_b$
has been depicted before \cite{Sel3}; there, the $\langle 32^3 \rangle$ phase
is squeezed out). Thence, by
lowering the surface couplings, the 3--band eventually
sticks at the surface. In fact, the same phenomenon
holds for wider odd films as well. The
transition line, separating the structures with one 1--band
from that with one 3--band and arising
from $(T=0, \kappa=1)$, terminates again
in a critical point for moderate and weak surface couplings. The point
moves to zero temperature as $J_s$ vanishes. Perhaps most
interestingly, mean--field theory suggests
that the $\langle 54 \rangle$ phase is
stable at intermediate temperatures, as shown in Fig. 4. Indeed, its
existence has been confirmed in Monte Carlo simulations for
equal surface, $J_s$, and bulk, $J_b$, couplings. The phase shows up below the
$\langle 41_04 \rangle$ phase which, in turn, becomes stable
at about the transition
temperature of the two--dimensional Ising model, like the
corresponding partially disordered phases for films with an
odd number of layers.

\begin{figure}
\centerline{\psfig{figure=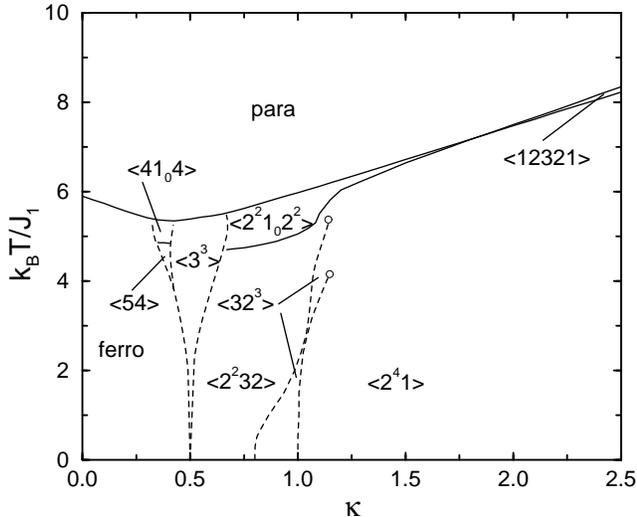,angle=270,width=3.3in}}  
\caption{Phase diagram of the ANNNI model for $L= 9$ at $J_s/J_b$= 0.95
using mean--field theory.}
\label{fig4}
\end{figure}

Finally, for $L=10$, seven distinct ordered phases have been
identified and located applying mean--field theory
for $J_s= J_b$ \cite{Sel3}. Close
to the transition to the
paramagnetic phase, $T_c$, the ferromagnetic, $\langle 5^2 \rangle$, 
$\langle 343 \rangle$, $\langle 23^22 \rangle$, and
$\langle 2^5 \rangle$ phases occur. The $\langle 43^2 \rangle$ phase
springs from the
multiphase point $(T=0, \kappa= 1/2)$. At intermediate temperatures, the
$\langle 64 \rangle$ phase has been detected in a tiny region of the phase
diagram. The phase persists, when varying the surface couplings. For
sufficiently small values of $J_s$, the
$\langle 32^23 \rangle$ phase is stabilized, at the
cost of the $\langle 23^22 \rangle$ phase, near the
multiphase point, reflecting again the gain of entropy due to easy
excitations of surface spins belonging to 3--bands for structures
comprising 3--bands and 2--bands.

\section{Summary}

The ANNNI model on thin films has been studied, using
mean--field theory, low--temperature analyses, and Monte Carlo
techniques. For each film thickness, varying the number of
layers from $L= 3$ to $L= 10$, distinct phase diagrams have
been determined, monitoring surface--induced features by
also changing the strength of the couplings in the surface
layers.

In the limit of infinite lattices, $L \longrightarrow \infty$, the
competing interactions of the ANNNI model lead to a phase diagram
with a rich variety of spatially modulated magnetic
structures. Signatures of some of these features, like the
sequence of commensurate phases springing from the multiphase
point, are already present in thin films, with the films displaying
generic and distinct features as well. In particular: (i) The ordered
phases occuring directly below the transition line to the
paramagnetic phase are alternatingly, as the ratio $\kappa= -J_2/J_1$
between the competing interactions of the model increases, symmetric
and antisymmetric about the center of the film, leading to
partially disordered phases with a paramagnetic
center plane for films with an odd number of layers. (ii) The phases
springing from the multiphase point may include structures consisting
not only of 2--bands and 3--bands, but also with a 4--band or
a 5--band. (iii) For $L$ odd, there is an additional transition line
near $\kappa= 1$, separating phases with 2--bands augmented
by one 3--band or one 1--band.

By modifying the surface couplings, additional interesting
phenomena are induced. Indeed, the range of stability of
the various phases as well as the type of phase transition may be
affected by varying the couplings. Entire transition lines may, in
fact, disappear. Perhaps most noticeable, 3--bands may
be forced to stick to the surface or move to the interior of
the film, thereby possibly forming new phases with a different
symmetry.

These findings may encourage experimental work
on thin films of magnets or alloys showing complicated spatial
orderings in the bulk. On the theoretical side, several
extensions of the present study seem to be promising, including
the effects of a larger number of layers, modified
interactions \cite{Grou,Mass,Mur,All}, and
magnetic bulk as well as surface fields, in order to possibly detect 
additional systematics in the intriguing phase diagrams of the
ANNNI model and variants.

\acknowledgments
We would like to thank A. G\"odecke for useful discussions. 


\end{document}